\documentclass[12pt]{article}
\usepackage[T1]{fontenc}
\usepackage[latin1]{inputenc}
\usepackage{geometry}
\usepackage{amsmath}
\usepackage{amssymb}

\makeatletter

\providecommand{\LyX}{L\kern-.1667em\lower.25em\hbox{Y}\kern-.125emX\@}

\usepackage{axodraw}
\makeatother

\makeatother

\begin{document}

\title{\hfill{}{\normalsize PCCF RI-00-19}\\
Supersymmetric penguin contributions to the decay \( b\rightarrow s\gamma  \)
with non-universal squarks masses}

\author{{\normalsize M.B. Causse}\thanks{
email: mbc@wanadoo.fr
} {\normalsize and J. Orloff}\thanks{
email: orloff@in2p3.fr
} {\normalsize }\\
{\normalsize Laboratoire de Physique Corpusculaire,}\\
{\normalsize Universit\'{e} Blaise Pascal, }\\
{\normalsize F-63177 Aubi\`{e}re Cedex}\normalsize }
\date{}
\maketitle
\begin{abstract}
We give explicit expressions for the amplitudes associated with the supersymmetric
(SUSY) contributions to the process: \( b\rightarrow s\gamma  \) in the context
of SUSY extensions of Standard Model (SM) with non-universal soft SUSY breaking
terms. From experimental data, we deduce limits on the squark mass insertions
obtained from different contributions (gluinos, neutralinos and charginos). 
\end{abstract}

\section{Introduction}

The rare \( B \) decays represent a good test for new physics beyond the SM
since they are not affected appreciably by uncertainties due to long distance
effects. Here, in the context of spontaneously broken minimal \( N=1 \) supergravity
\cite{1}, we study penguin diagrams with gluinos, neutralinos and charginos,
which are responsible \( \Delta S=1 \) radiative mesonic decays. In particular,
we study the \( b\rightarrow s\gamma  \) decay \cite{2} that gives the most
stringent lower bounds on the average squark mass. We know that in generic msugra
models \cite{3}, the soft universal breaking terms lead to a high degeneracy
in the sfermionic sector.  Flavor changing neutral current (FCNC) tests play
an important role to constrain the SUSY mass spectrum. 

We thus consider the SUSY extensions of the SM with non-universal soft breaking
terms \cite{5}. We shall use the mass insertion method by which it is possible
to obtain a set of upper bounds on the off-diagonal terms (\( \Delta  \)) in
the sfermion mass matrices (the mass terms relating sfermion of the same electric
charge but different flavor). Obviously the mass insertion method offers the
major advantage that one does not need the full diagonalization of the sfermion
mass matrices. Then only a small number of effective parameters (\( \delta  \))
summarize the effects. We have applied this method to the gluinos,  neutralinos
and charginos contributions to the decay \( b\rightarrow s\gamma  \), the charginos
contribution being original. From experimental limits, we have then derived
upper bounds on the off-diagonal terms in the sfermion mass matrices (for squarks
down and up). 

This paper is organized as follows. In section 2 we present generalities on
the minimal SUSY models and the mass insertion method. In section 3, we give
the explicit expressions of the amplitudes associated with the gluinos, neutralinos
and charginos contributions to the decay: \( b\rightarrow s\gamma  \). Finally
in section 4, we find explicit expressions for the branching ratio \( BR(b\rightarrow s\gamma ) \)
and the upper bounds obtained for the off-diagonal terms (\( \Delta  \)) in
the squark mass matrices. In the annex, we recall the analytic expressions for
the Feynman integrals which arise in the evaluation of these amplitudes.

\section{Generalities on the minimal SUSY models and the mass insertion method. }

The minimal supersymmetric standard model (MSSM), obtained by supersymmetrizing
the SM field contents and allowing for all possible soft susy-breaking terms,
contains a huge number of free parameters. In this note, we concentrate on a
specific set of models where these soft-breaking terms are close to the msugra
universality. 

By minimal supergravity models \cite{3} (msugra) we below mean, the low energy
limit of spontaneously broken \( N=1 \) supergravity theories which supersymmetrize
the SM and present the following two features : 

\begin{itemize}
\item \( R \)-parity is implemented so that no baryon and/or lepton number violating
terms appear in the superpotential, 
\item the Kälher metric is flat, i.e. all the scalar kinetic terms are canonical.
 
\end{itemize}
These features bring about new sources of FCNC (flavor changing neutral current)
effects. The experimental limits on \( B \) meson physics and in particular
\( b\rightarrow s\gamma  \), constitute interesting FCNC tests for these MSS
Models, typically requiring squark masses of the same electric charge to be
relatively degenerate,  i.e. their mass difference must be smaller than their
average value.  Briefly, we review the major ingredients which give rise to
this new source of FCNC.  

The low energy Lagrangian consists of :

\begin{itemize}
\item The superpotential of the \( N=1 \) globally supersymmetric SM:
\end{itemize}
\begin{equation}
\label{superpot}
W=h_{u}QH_{1}u^{c}+h_{d}QH_{2}d^{c}+h_{L}LH_{2}e^{c}+\mu H_{2}H_{1}
\end{equation}

\begin{itemize}
\item The scalar part of SUSY soft breaking terms for the minimal \( N=1 \) supergravity
theories : \begin{equation}
\label{softs}
L^{scal}_{soft}=m^{2}\times \sum _{i=scalar}\left| \varphi _{i}\right| ^{2}+\left[ Am\left( h_{u}\widetilde{Q}H_{1}\widetilde{u^{c}}+h_{u}\widetilde{Q}H_{2}\widetilde{d^{c}}+h_{L}\widetilde{L}H_{2}\widetilde{e^{c}}\right) +Bm\mu H_{2}H_{1}+h.c\right] 
\end{equation}
where \( A \) and \( B \) are two dimensionless free parameters of the trilinear
and bilinear scalar contributions; \( m \) denote the scale of the low energy
SUSY breaking. 
\end{itemize}
From equations (\ref{superpot}) and (\ref{softs}), we obtain the squark down
\( 6\times 6 \) mass matrix \( (Q=-\frac{1}{3}) \)

\begin{equation}
\label{matricesd}
M_{\widetilde{d}\widetilde{d^{*}}}^{2}=\left[ \begin{array}{cc}
m_{\widetilde{d_{L}}\widetilde{d_{L}^{*}}}^{2} & m_{\widetilde{d_{L}}\widetilde{d_{L}^{c}}}^{2}\\
m_{\widetilde{d_{L}^{*}}\widetilde{d_{L}^{c*}}}^{2} & m_{\widetilde{d_{L}^{c}}\widetilde{d_{L}^{c*}}}^{2}
\end{array}\right] 
\end{equation}
Where\begin{equation}
\label{massesd}
m_{\widetilde{d_{L}}\widetilde{d^{*}_{L}}}^{2}=m_{\widetilde{d_{L}^{c}}\widetilde{d_{L}^{c*}}}^{2}=m_{d}m^{+}_{d}+m^{2}\times 1
\end{equation}
and

\begin{equation}
\label{remassesd}
m_{\widetilde{d_{L}}\widetilde{d^{c}_{L}}}^{2}=Amm_{d}+\mu m_{d}\langle H_{1}\rangle /\langle H_{2}\rangle 
\end{equation}
 with \( m_{d}= \) \( 3\times 3 \) quarks \( D \) mass matrix and \( e_{D}=-\frac{1}{3} \)
is the electric charge. 

At this stage, it is clear that the \( d_{L}-\widetilde{d_{L}^{+}}-\widetilde{g} \)
coupling cannot lead to flavor changes (FC). Indeed, if we diagonalize \( m_{d}m_{d}^{+} \),
we diagonalize at the same time \( m_{\widetilde{d_{L}}\widetilde{d^{*}_{L}}}^{2} \).
However, this is no longer true if we renormalize \( m_{\widetilde{d_{L}}\widetilde{d^{*}_{L}}}^{2} \):
its value stays (\ref{massesd}) at the high scale, but its evolution down to
the \( M_{W} \) scale  cannot be diagonal due to the \( h_{u}QH_{1}u^{c} \)
term in the superpotential (\ref{superpot}). Hence the \( 3\times 3 \) mass
matrix of \( m_{\widetilde{d}\widetilde{d^{*}}}^{2} \) renormalized to the
\( M_{W} \) scale, reads: \begin{equation}
\label{newsd}
m_{\widetilde{d_{L}}\widetilde{d^{*}_{L}}}^{2}(q^{2}=M^{2}_{w})=m_{d}m_{d^{+}}+m^{2}\times 1+cm_{u}m^{+}_{u}
\end{equation}
 Where the coefficient \( c \) can be computed by solving the set of renormalization
group equations for the evolution of the SUSY quantities. From equ. (\ref{newsd})
we see that the simultaneous diagonalization of \( m_{\widetilde{d_{L}}\widetilde{d^{*}_{L}}}^{2} \)
and \( m_{d}m_{d^{+}} \) is no longer possible due to the presence of the \( cm_{u}m^{+}_{u} \)
term. The flavor change is proportional to \( c \) and to the usual (CKM) angles.
In a basis where \( d_{L}-\widetilde{d_{L}^{+}}-\widetilde{g} \) couplings
are flavor diagonal, the flavor mixing occurs in the squark propagators.  The
above remark can be summarized in the following schematic way:

\begin{equation}
\begin{picture}(80,30)(0,0)
\DashLine(0,20)(80,20){3}
\Line(35,15)(45,25) \Line(35,25)(45,15)
\Text(40,13)[tc]{$\Delta_{LL}^{ij}$}
\Text(20,22)[bc]{$\tilde d_{iL}$}
\Text(60,22)[bc]{$\tilde d_{jL}$}
\end{picture}
\qquad\rightarrow\qquad \Delta_{LL}^{ij}=c(V.[m^{diag}_u]^2.V^\dagger)_{ij}
\end{equation}\\
For the \( L\rightarrow R \) transitions,  we have:

\begin{equation}
\begin{picture}(80,30)(0,0)
\DashLine(0,20)(80,20){3}
\Line(35,15)(45,25) \Line(35,25)(45,15)
\Text(40,13)[tc]{$\Delta_{LR}^{ij}$}
\Text(20,22)[bc]{$\tilde d_{iL}$}
\Text(60,22)[bc]{$\tilde d^c_{jL}$}
\end{picture} 
\qquad\rightarrow\qquad \Delta_{LR}^{ij}=\Delta_{LR}^{bs}
\end{equation}\\
The quantities \( \Delta _{ij} \) are mass insertions connecting flavors \( i \)
and \( j \) along a sfermion propagator and the indices \( L,R \) refer to
the helicity of the fermion partners.  There are three types of sfermions mixing:
\( \Delta _{LL} \), \( \Delta _{RR} \) and \( \Delta _{LR} \). In the MSSM
case with universal soft SUSY breaking (msugra), there exists a kind of hierarchy
among mass insertions that is \( (\Delta _{LL})_{ij}\gg (\Delta _{LR})_{ij}\gg (\Delta _{RR})_{ij} \).
This is no longer true if flavor changing is produced by another kind of {}``initial{}''
conditions. Then, generally, nothing can be said about the hierarchy of these
3 contributions. In that case, one needs a model-independent parameterization
of the flavor changing (FC) and CP quantities in SUSY to test variants
of the universal MSSM. The chosen parameterization is the mass insertion approximation
\cite{6}-\cite{8}. It concerns the most peculiar source of FCNC SUSY contributions
that do not arise from the mere supersymmetrization of the FCNC in the SM. They
originate from the FC coupling of gluinos, neutralinos and charginos to fermions
and sfermions. One chooses a basis for fermions and sfermions states where all
couplings of these particles to gauginos are flavor diagonal, while the FC originates
from non-diagonal sfermion mass terms in propagators. Denoting by \( \Delta  \)
the off-diagonal terms in the sfermions mass matrix (i.e. the mass terms relating
sfermions of the same electric charge,  but different flavor), the sfermion
propagators can be expanded as a series of \( \delta =\Delta /\tilde{m}^{2} \)
where \( \widetilde{m} \) is an average sfermions mass and a typical scale
of the SUSY breaking. As long as the ratio of non-diagonal entries (\( \Delta  \))
to the average squark mass is a small parameter\cite{5}, the first term in
the expansion, obtained from the non-diagonal insertion of mass between two
diagonal squark propagators, represents an reasonable approximation. This method
has the advantage that one does not need the full diagonalization of the sfermion
mass matrices. So, from the FCNC experimental data we may derive upper bounds
on the different \( \delta  \)'s.  

In the following section, we give the explicit expression of the amplitudes
associated to the gluinos, neutralinos and charginos contributions to the decay:
\( b\rightarrow s\gamma  \) \cite{2},\cite{10},\cite{11}. 

\section{\label{Absg}Amplitudes contributing to the decay \protect\( b\rightarrow s\gamma \protect \)}

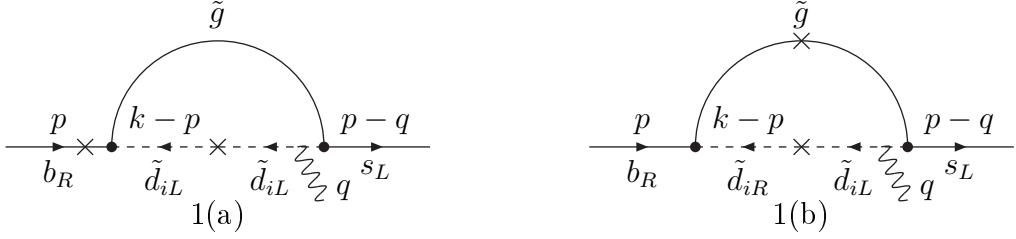
\begin{figure}
{\par\centering \begin{picture}(160,80)(0,0)
\ArrowLine(0,20)(40,20)\Text(20,15)[tc]{\(b_{R}\)}\Text(20,25)[bc]{\(p\)}
\Vertex(40,20){2}
\DashArrowLine(80,20)(40,20){3}\Text(60,15)[tc]{\(\tilde {d}_{iL}\)}\Text(60,25)[bc]{\(k-p\)}
\Line(77,17)(83,23)\Line(77,23)(83,17)
\Text(80,0)[tc]{\ref{f1}(a)}
\DashArrowLine(120,20)(80,20){3}\Text(100,15)[tc]{\(\tilde {d}_{iL}\)}
\Vertex(120,20){2}
\ArrowLine(120,20)(160,20)\Text(140,15)[tc]{\(s_{L}\)}\Text(140,25)[bc]{\(p-q\)}
\CArc(80,20)(40,0,180)
\Line(27,17)(33,23)\Line(27,23)(33,17) 
\Text(80,65)[bc]{\(\tilde {g}\)}
\Photon(110,20)(120,0){3}{4}\Text(125,0)[bl]{\(q\)}
\end{picture} \hskip2cm \begin{picture}(160,80)(0,0)
\ArrowLine(0,20)(40,20)\Text(20,15)[tc]{\(b_{R}\)}\Text(20,25)[bc]{\(p\)}
\Vertex(40,20){2}
\DashArrowLine(80,20)(40,20){3}\Text(60,15)[tc]{\(\tilde {d}_{iR}\)}\Text(60,25)[bc]{\(k-p\)}
\Line(77,17)(83,23)\Line(77,23)(83,17)
\Text(80,0)[tc]{\ref{f1}(b)}
\DashArrowLine(120,20)(80,20){3}\Text(100,15)[tc]{\(\tilde {d}_{iL}\)}
\Vertex(120,20){2}
\ArrowLine(120,20)(160,20)\Text(140,15)[tc]{\(s_{L}\)}\Text(140,25)[bc]{\(p-q\)}
\CArc(80,20)(40,0,180)
\Line(77,57)(83,63)\Line(77,63)(83,57) \Text(80,65)[bc]{\(\tilde {g}\)}
\Photon(110,20)(120,0){3}{4}\Text(125,0)[bl]{\(q\)}
\end{picture}\par}

\caption{\label{f1}Gluino contribution  (pengluinos).}
\end{figure}

\begin{figure}
{\par\centering \begin{picture}(160,80)(0,0)
\ArrowLine(0,20)(40,20)\Text(20,15)[tc]{\(b_{R}\)}\Text(20,25)[bc]{\(p\)}
\Vertex(40,20){2}
\DashArrowLine(80,20)(40,20){3}\Text(60,15)[tc]{\(\tilde {d}_{iL}\)}\Text(60,25)[bc]{\(k-p\)}
\Line(77,17)(83,23)\Line(77,23)(83,17)
\Text(80,0)[tc]{\ref{f2}(a)}
\DashArrowLine(120,20)(80,20){3}\Text(100,15)[tc]{\(\tilde {d}_{iL}\)}
\Vertex(120,20){2}
\ArrowLine(120,20)(160,20)\Text(140,15)[tc]{\(s_{L}\)}\Text(140,25)[bc]{\(p-q\)}
\CArc(80,20)(40,0,180)
\Line(27,17)(33,23)\Line(27,23)(33,17) 
\Text(80,65)[bc]{\(\chi^{0}_{j}\)}
\Photon(110,20)(120,0){3}{4}\Text(125,0)[bl]{\(q\)}
\end{picture} \hskip2cm \begin{picture}(160,80)(0,0)
\ArrowLine(0,20)(40,20)\Text(20,15)[tc]{\(b_{R}\)}\Text(20,25)[bc]{\(p\)}
\Vertex(40,20){2}
\DashArrowLine(80,20)(40,20){3}\Text(60,15)[tc]{\(\tilde {d}_{R}\)}\Text(60,25)[bc]{\(k-p\)}
\Line(77,17)(83,23)\Line(77,23)(83,17)
\Text(80,0)[tc]{\ref{f2}(b)}
\DashArrowLine(120,20)(80,20){3}\Text(100,15)[tc]{\(\tilde {d}_{iL}\)}
\Vertex(120,20){2}
\ArrowLine(120,20)(160,20)\Text(140,15)[tc]{\(s_{L}\)}\Text(140,25)[bc]{\(p-q\)}
\CArc(80,20)(40,0,180)
\Line(77,57)(83,63)\Line(77,63)(83,57) \Text(80,65)[bc]{\(\chi^{0}_{j}\)}
\Photon(110,20)(120,0){3}{4}\Text(125,0)[bl]{\(q\)}
\end{picture}\par}

\caption{\label{f2}Neutralino contributions (penneutralinos).}
\end{figure}
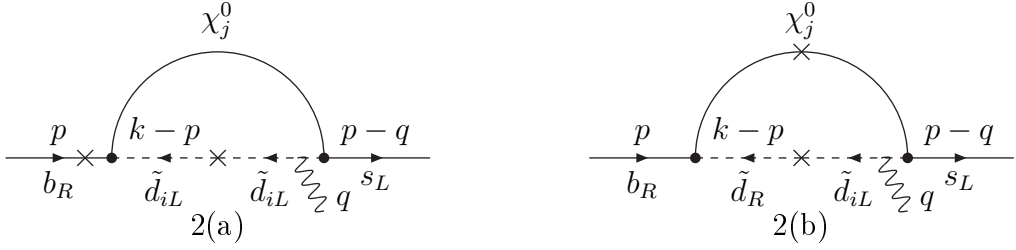

The MSSM Feynman rules used for the calculation of the amplitudes can be found
in the reference \cite{9}. The calculation of these amplitudes is done with
these Feynman rules and the mass insertion approximation. Supersymmetric penguin
diagrams contributing to the decay \( b\rightarrow s\gamma  \) are:

\begin{itemize}
\item gluinos (pengluinos): figures \ref{f1}(a) and \ref{f1}(b)
\item neutralinos (penneutralinos): figures \ref{f2}(a) and \ref{f2}(b) 
\item charginos (pencharginos): figures \ref{f3}(a) and \ref{f3}(b); figures \ref{f4}(a)
and \ref{f4}(b); figures \ref{f5}(a) and \ref{f5}(b). 
\end{itemize}
These diagrams induce the effective operator \( O_{LR}=m_{b}\varepsilon _{\mu }(q)\overline{s(p-q)}\sigma ^{\mu \nu }q_{\nu }P_{R}b(p) \),
\( q \) is the outgoing momentum of the photon. 

Now we can give the explicit expressions of the amplitudes associated with the
supersymmetric penguin diagrams. 

\subsection{The pengluinos}

From the diagram illustrated in figure \ref{f1}(a), we obtain 

\begin{equation}
\label{ampgluino}
T^{'}_{\widetilde{g_{LL}}}=e_{D}C_{2}(R)\alpha _{s}\frac{1}{\sqrt{\pi }}\sqrt{\alpha }\frac{\delta _{LL}}{M^{2}_{D}}\varepsilon _{\mu }(q)\overline{s(p-q)}\sigma ^{\mu \nu }q_{\nu }P_{R}[pH(X_{\widetilde{g}})+qH(X_{\widetilde{g}})]b(p)
\end{equation}
where \( D_{i} \) with \( i=1.....6 \) are squark down mass eigenstates, \( e_{D} \)
is the electric charge of the squark D; \( X_{\widetilde{g}}=\frac{M^{2}_{\widetilde{g}}}{M^{2}_{D}} \)
and the function \( H(X_{\widetilde{g}}) \) is given in annex; \( \delta _{LL} \)
is the mass insertion connecting flavors \( b \) and \( s \) with the helicity
\( L \): 
\begin{equation}\label{delta_LL}
\delta _{LL}=\sum
^{6}_{i=1}\frac{(M^{2}_{Di}-M_{D}^{2})Z^{si*}_{D}Z^{bi}_{D}}{M^{2}_{D}}=\frac{\Delta
  _{LL}}{M^{2}_{D}}
\end{equation}
 in which \( Z_{D} \) is a mixing matrix defined by: \[
diag(M_{D1}^{2}......M_{D6}^{2})=Z_{D}^{+}\left( \begin{array}{cc}
M_{LL}^{2} & M_{LR}^{2+}\\
M_{LR}^{2} & M_{RR}^{2}
\end{array}\right) Z_{D}\, \, ;\]
 \( M^{2}_{D} \) is the average squark down mass and \( C_{2}(R)=\frac{4}{3} \)
(fundamental representation ).So the \( T^{'}_{\widetilde{g_{LL}}} \) expression
becomes: \[
T^{'}_{\widetilde{g_{LL}}}=e_{D}C_{2}(R)\alpha _{s}\frac{\delta _{LL}}{M^{2}_{D}\sqrt{\pi }}\sqrt{\alpha }\varepsilon _{\mu }(q)\overline{s(p-q)}\sigma ^{\mu \nu }q_{\nu }P_{R}[pH(X_{\widetilde{g}})+qH(X_{\widetilde{g}})]b(p)\]

From the diagram drawn in figure 1(b) we have \[
T^{'}_{\widetilde{g_{LR}}}=e_{D}C_{2}(R)\alpha _{s}\frac{M_{\widetilde{g}}\delta _{LR}}{M^{2}_{D}\sqrt{\pi }}\sqrt{\alpha }M_{1}(X_{\widetilde{g}})\varepsilon _{\mu }(q)\overline{s(p-q)}\sigma ^{\mu \nu }q_{\nu }P_{R}b(p)\]
 with\[
\delta _{LR}=\sum ^{6}_{i=1}\frac{(M_{Di}^{2}-M_{D}^{2})Z^{si*}_{D}Z^{(b+3)i}_{D}}{M^{2}_{D}}=\frac{\Delta _{LR}}{M^{2}_{D}}\]
 Thanks to the experimental limits,  it will be possible to put upper bounds
on the different \( \delta  \)'s, that is on the non-diagonal terms in the
sfermion mass matrix. 

\subsection{The penneutralinos }

The penneutralinos are illustrated in figures \ref{f2}(a) and \ref{f2}(b). 

\begin{itemize}
\item for the diagram (a)\[
T^{'}_{\chi ^{0}_{LLj}}=e_{D}\alpha _{w}\frac{\delta _{LL}}{2\cos ^{2}(\theta _{w})M^{2}_{D}\sqrt{\pi }}\sqrt{\alpha }(z_{L\chi ^{0}_{j}})\overline{s(p-q)}\varepsilon _{\mu }(q)\sigma ^{\mu \nu }q_{\nu }P_{R}[pH(X_{0j})-qH(X_{0j})/3]b(p)\]
\( j=1...4 \) the four neutralinos indices,  \( X_{0j}=\frac{M^{2}_{\chi ^{0}_{j}}}{M^{2}_{D}} \)
with \( M_{\chi ^{0}_{j}} \) the neutralino mass.  and \[
z_{L\chi ^{0}_{j}}=\left| \frac{1}{3}Z^{1j}_{N}\sin \theta _{w}-Z^{2j}_{N}\cos \theta _{w}\right| ^{2}\]
 Clearly, \( z_{L\chi ^{0}_{j}} \) is less or equal to \( 1 \). In msugra,
for example, we will have: \( z_{L\chi ^{0}_{1}}\approx \frac{\sin ^{2}(\theta _{w})}{9} \)
because \( Z^{11}_{N}\approx 1 \) for the lightest neutralino ( bino-like),
and \( z_{L\chi ^{0}_{2}}\cong 0.8 \) for \( Z^{22}_{N}\approx 1 \) 
\item for the diagram (b) \[
T^{'}_{\chi ^{0}_{RLj}}=-e_{D}\alpha _{w}\frac{\sin (\theta _{w})\delta _{LR}M_{\chi _{j}^{0}}}{3\cos ^{2}(\theta _{w})M^{2}_{D}\sqrt{\pi }}\sqrt{\alpha }(z_{Rx^{0}_{j}})M_{1}(X_{0j})\overline{s(p-q)}\varepsilon _{\mu }(q)\sigma ^{\mu \nu }q_{\nu }P_{R}b(p)\]
 with\[
z_{R\chi ^{0}_{j}}=(\frac{1}{3}Z^{1j*}_{N}\sin \theta _{w}-Z_{N}^{2j*}\cos \theta _{w})Z^{1j*}_{N}\]
 where, in msugra, we can have \( z_{R\chi ^{0}_{2}}\approx 1 \). 
\begin{figure}
{\par\centering \begin{picture}(160,80)(0,0)
\ArrowLine(0,20)(40,20)\Text(20,15)[tc]{\(b_{R}\)}\Text(20,25)[bc]{\(p\)}
\Vertex(40,20){2}
\DashArrowLine(80,20)(40,20){3}\Text(60,15)[tc]{\(\tilde {u}_{iL}\)}\Text(60,25)[bc]{\(k-p\)}
\Line(77,17)(83,23)\Line(77,23)(83,17)
\Text(80,0)[tc]{\ref{f3}(a)}
\DashArrowLine(120,20)(80,20){3}\Text(100,15)[tc]{\(\tilde {u}_{iL}\)}
\Vertex(120,20){2}
\ArrowLine(120,20)(160,20)\Text(140,15)[tc]{\(s_{L}\)}\Text(140,25)[bc]{\(p-q\)}
\CArc(80,20)(40,0,180)
\Line(77,57)(83,63)\Line(77,63)(83,57) \Text(80,65)[bc]{\(\chi^{-}_{j}\)}
\Photon(110,20)(120,0){3}{4}\Text(125,0)[bl]{\(q\)}
\end{picture} \hskip2cm \begin{picture}(160,80)(0,0)
\ArrowLine(0,20)(40,20)\Text(20,15)[tc]{\(b_{R}\)}\Text(20,25)[bc]{\(p\)}
\Vertex(40,20){2}
\DashArrowLine(80,20)(40,20){3}\Text(60,15)[tc]{\(\tilde {u}_{iL}\)}\Text(60,25)[bc]{\(k-p\)}
\Line(77,17)(83,23)\Line(77,23)(83,17)
\Text(80,0)[tc]{\ref{f3}(b)}
\DashArrowLine(120,20)(80,20){3}\Text(100,15)[tc]{\(\tilde {u}_{iL}\)}
\Vertex(120,20){2}
\ArrowLine(120,20)(160,20)\Text(140,15)[tc]{\(s_{L}\)}\Text(140,25)[bc]{\(p-q\)}
\CArc(80,20)(40,0,180)
\Line(27,17)(33,23)\Line(27,23)(33,17) 
\Text(80,65)[bc]{\(\chi^{-}_{j}\)}
\Photon(110,20)(120,0){3}{4}\Text(125,0)[bl]{\(q\)}
\end{picture}\par}

\caption{\label{f3}Chargino contributions with photon coupling to up squark and mass
insertion \protect\( LL\protect \).}
\end{figure}
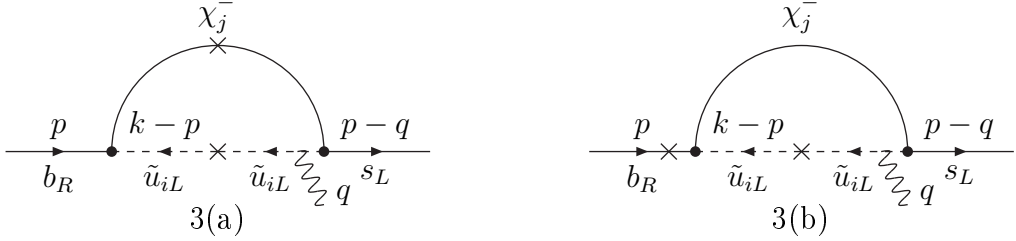

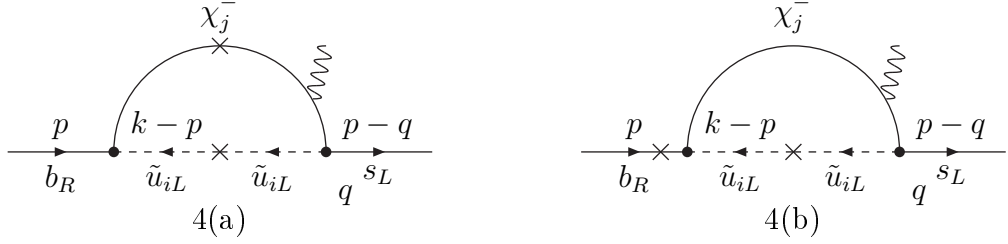
\begin{figure}
{\par\centering \begin{picture}(160,80)(0,0)
\ArrowLine(0,20)(40,20)\Text(20,15)[tc]{\(b_{R}\)}\Text(20,25)[bc]{\(p\)}
\Vertex(40,20){2}
\DashArrowLine(80,20)(40,20){3}\Text(60,15)[tc]{\(\tilde {u}_{iL}\)}\Text(60,25)[bc]{\(k-p\)}
\Line(77,17)(83,23)\Line(77,23)(83,17)
\Text(80,0)[tc]{\ref{f4}(a)}
\DashArrowLine(120,20)(80,20){3}\Text(100,15)[tc]{\(\tilde {u}_{iL}\)}
\Vertex(120,20){2}
\ArrowLine(120,20)(160,20)\Text(140,15)[tc]{\(s_{L}\)}\Text(140,25)[bc]{\(p-q\)}
\CArc(80,20)(40,0,180)
\Line(77,57)(83,63)\Line(77,63)(83,57) \Text(80,65)[bc]{\(\chi^{-}_{j}\)}
\Photon(115,40)(120,60){-3}{4}\Text(125,0)[bl]{\(q\)}
\end{picture}\hskip2cm \begin{picture}(160,80)(0,0)
\ArrowLine(0,20)(40,20)\Text(20,15)[tc]{\(b_{R}\)}\Text(20,25)[bc]{\(p\)}
\Vertex(40,20){2}
\DashArrowLine(80,20)(40,20){3}\Text(60,15)[tc]{\(\tilde {u}_{iL}\)}\Text(60,25)[bc]{\(k-p\)}
\Line(77,17)(83,23)\Line(77,23)(83,17)
\Text(80,0)[tc]{\ref{f4}(b)}
\DashArrowLine(120,20)(80,20){3}\Text(100,15)[tc]{\(\tilde {u}_{iL}\)}
\Vertex(120,20){2}
\ArrowLine(120,20)(160,20)\Text(140,15)[tc]{\(s_{L}\)}\Text(140,25)[bc]{\(p-q\)}
\CArc(80,20)(40,0,180)
\Line(27,17)(33,23)\Line(27,23)(33,17) 
\Text(80,65)[bc]{\(\chi^{-}_{j}\)}
\Photon(115,40)(120,60){-3}{4}\Text(125,0)[bl]{\(q\)}
\end{picture}\par}

\caption{\label{f4}Chargino contributions with photon coupling to the chargino and
mass insertion \protect\( LL\protect \).}
\end{figure}

\end{itemize}

\subsection{The pencharginos}

Due to the photon-chargino coupling, there are \( 6 \) diagrams. For the mass
insertion \( LL \), 4 diagrams illustrated in figures \ref{f3}(a) and \ref{f3}(b)
and figures \ref{f4}(a-b) contribute. When the helicity flip is realized in
the quark \( b \) external line, only the wino component of chargino is concerned
in the calculation of the amplitude (diagrams b).But in the case where the helicity
flip is realized on the chargino line the higgsino components are taken into
account(diagrams a). From the penchargino diagrams: figures \ref{f3}(a) and
\ref{f3}(b), where the photon is coupled to the squarks, we obtain: 

\begin{itemize}
\item for the diagram (a)\[
T^{'}_{\chi ^{-}_{LLj}}=e_{u}\alpha _{w}\frac{m_{b}\delta _{LL\chi _{j}}M_{\chi ^{-}_{j}}Z^{+*}_{1j}Z^{-*}_{2j}}{M_{w}\cos (\beta )M^{2}_{U}\sqrt{2\pi }}\sqrt{\alpha }M_{1}(X_{j})\overline{s(p-q)}\varepsilon _{\mu }(q)\sigma ^{\mu \nu }q_{\nu }P_{R}b(p)\]
 where \( j=1,2 \) are the two chargino states, \( X_{j}=\frac{M^{2}_{\chi _{j}^{-}}}{M^{2}_{U}} \)
and \( M_{\chi ^{-}_{j}} \) is the\( j \) chargino mass, \( e_{u}=2/3 \)
is the squarks up electric charge,  \( M_{U} \) the squark up average mass.
And
\begin{equation}\label{delta_LLXj}
\delta _{LL\chi _{j}}=\sum ^{6}_{i=1}\sum ^{3}_{J=1}\sum ^{3}_{K=1}(1-\delta
_{JK})\frac{(M_{Ui}^{2}-M_{U}^{2})}{M^{2}_{U}}Z^{Ki}_{U}V_{sK}Z^{Ji*}_{U}V_{bJ}^{*}
\end{equation}
 in which \( J \) and \( K \) run over the 3 generations of squarks and \( U_{i} \)
with \( i=1,\ldots 6 \) are up squarks mass eigenstates.  As in \( \delta _{LL} \),
\( \delta _{LL\chi j} \) contains squark mixing factors $Z$,  but in addition,
there are some known Cabbibo quarks mixing factors ({\it e.g.} \( V_{bc} \)). 
\item for the diagram (b) \[
T^{'}_{\chi ^{-}_{LLj}}=e_{u}\alpha _{w}\frac{\delta _{LL\chi_{j}}Z^{+*}_{1j}Z^{+}_{1j}}{M^{2}_{U}\sqrt{\pi }}\sqrt{\alpha }\overline{s(p-q)}\varepsilon _{\mu }(q)\sigma ^{\mu \nu }q_{\nu }P_{R}[pH(X_{j})+qH(X_{j})/3]b(p)\]
 In msugra for the lightest chargino,  we have \( Z^{+*}_{1j}\approx 1 \) when
\( Z^{-*}_{2j}\approx 0 \); in such case, we remark that only the diagram (b)
( wino component of chargino) contribute to the amplitude. 
\end{itemize}
The pencharginos illustrated in figures \ref{f4}(a) and \ref{f4}(b), where
the photon is coupled to the chargino, give: 

\begin{itemize}
\item for diagram (a): \[
T^{'}_{\chi ^{-}_{LLj}}=-\alpha _{w}\frac{m_{b}\delta _{LL\chi _{j}}M_{\chi ^{-}_{j}}Z^{+*}_{1j}Z^{-*}_{2j}}{M^{2}_{w}\cos (\beta )M^{2}_{U}\sqrt{2\pi }}\sqrt{\alpha }F(X_{j})\overline{s(p-q)}\varepsilon _{\mu }(q)\sigma ^{\mu \nu }q_{\nu }P_{R}b(p)\]

\item for the diagram(b) \[
T^{'}_{\chi ^{-}_{LLj}}=\alpha _{w}\frac{\delta _{LL\chi _{j}}Z^{+*}_{1j}Z^{+}_{1j}}{M^{2}_{U}\sqrt{\pi }}\sqrt{\alpha }\overline{s(p-q)}\varepsilon _{\mu }(q)\sigma ^{\mu \nu }q_{\nu }P_{R}[pG(X_{j})+qG(X_{j})/2]b(p)\]
 We define the fraction of gaugino in the chargino \( j \) by: \( z_{\chi Lj}=Z_{1j}^{+*}Z_{1j}^{+} \).
the \( F(X_{j}) \) and \( G(X_{j}) \) are given in annex \ref{Annexe}. As
above,  in msugra, only the diagram (b) will contribute to the amplitude.
\end{itemize}

\begin{figure}
{\par\centering \begin{picture}(160,80)(0,0)
\ArrowLine(0,20)(40,20)\Text(20,15)[tc]{\(b_{R}\)}\Text(20,25)[bc]{\(p\)}
\Vertex(40,20){2}
\DashArrowLine(80,20)(40,20){3}\Text(60,15)[tc]{\(\tilde {u}_{iL}\)}\Text(60,25)[bc]{\(k-p\)}
\Line(77,17)(83,23)\Line(77,23)(83,17)
\Text(80,0)[tc]{\ref{f5}(a)}
\DashArrowLine(120,20)(80,20){3}\Text(100,15)[tc]{\(\tilde {u}_{iR}\)}
\Vertex(120,20){2}
\ArrowLine(120,20)(160,20)\Text(140,15)[tc]{\(s_{L}\)}\Text(140,25)[bc]{\(p-q\)}
\CArc(80,20)(40,0,180)
\Line(77,57)(83,63)\Line(77,63)(83,57) \Text(80,65)[bc]{\(\chi^{-}_{j}\)}
\Photon(110,20)(120,0){3}{4}\Text(125,0)[bl]{\(q\)}
\end{picture} \hskip2cm \begin{picture}(160,80)(0,0)
\ArrowLine(0,20)(40,20)\Text(20,15)[tc]{\(b_{R}\)}\Text(20,25)[bc]{\(p\)}
\Vertex(40,20){2}
\DashArrowLine(80,20)(40,20){3}\Text(60,15)[tc]{\(\tilde {u}_{iL}\)}\Text(60,25)[bc]{\(k-p\)}
\Line(77,17)(83,23)\Line(77,23)(83,17)
\Text(80,0)[tc]{\ref{f5}(b)}
\DashArrowLine(120,20)(80,20){3}\Text(100,15)[tc]{\(\tilde {u}_{iR}\)}
\Vertex(120,20){2}
\ArrowLine(120,20)(160,20)\Text(140,15)[tc]{\(s_{L}\)}\Text(140,25)[bc]{\(p-q\)}
\CArc(80,20)(40,0,180)
\Line(77,57)(83,63)\Line(77,63)(83,57) \Text(80,65)[bc]{\(\chi^{-}_{j}\)}
\Photon(115,40)(120,60){-3}{4}\Text(125,0)[bl]{\(q\)}
\end{picture}\par}

\caption{\label{f5}Chargino contributions with mass insertion \protect\( LR\protect \).}
\end{figure}
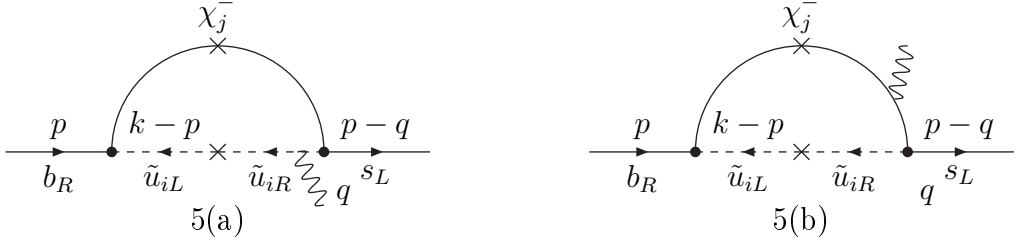
The \( LR \) mass insertion for the chargino contribution are illustrated in
figures \ref{f5}(a) and \ref{f5}(b). Only the higgsino components of the chargino
contributes to the amplitude. Therefore, due to the \( s_{L}-\chi ^{-}_{j}-U_{iR} \)
coupling giving a factor \( U^{J}Z^{(J+3)i}_{U}Z^{+*}_{2j}P_{R}V_{sJ} \) where
\( U^{J} \) is a Yukawa coupling proportional to the associated quark mass, the
top quark contribution overwhelms the up and charm ones. 

From diagram (a) in figure 5 we thus obtain: \[
T^{'}_{\chi ^{-}_{LRj}}=-e_{u}\alpha _{w}\frac{m_{t}m_{b}\delta _{LR\chi _{j}}z_{Rj}M_{\chi ^{-}_{j}}}{M^{2}_{W}\cos (\beta )\sin (\beta )M^{2}_{U}\sqrt{\pi }}\sqrt{\alpha }M_{1}(X_{j})\overline{s(p-q)}\varepsilon _{\mu }(q\sigma ^{\mu \nu }q_{\nu }P_{R}b(p)\]
 where\footnote{It would be desirable to pull out CKM factors from $\delta$'s,
   so that they only reflect squark properties. While this is arguably possible for
   $\delta_{LR\chi}$, it requires non-trivial assumptions for
   $\delta_{LL\chi}$, which is why we kept CKM factors in the definition of both.}
 \begin{equation}\label{delta_LRXj}
\delta _{LR\chi _{j}}=\sum ^{6}_{i=1}\frac{(M_{Ui}^{2}-M_{U}^{2})}{M^{2}_{U}}Z^{(t+3)i}_{U}V_{st}Z^{ti*}_{U}V^{*}_{bt}
\end{equation}
and for diagram (b) : \[
T^{'}_{\chi ^{-}_{LRj}}=\alpha _{w}\frac{m_{t}m_{b\, }\delta _{LR\chi _{j}\, }z_{Rj\, }M_{\chi ^{-}_{j}}}{2M^{2}_{W}\cos (\beta )\sin (\beta )M^{2}_{U}\sqrt{\pi }}\sqrt{\alpha }F(X_{j})\overline{s(p-q)}\varepsilon _{\mu }(q)\sigma ^{\mu \nu }q_{\nu }P_{R}b(p)\]
 with: \( z_{Rj}=Z^{+*}_{2j}Z^{-*}_{2j} \),  the fraction of higgsino in the
chargino \( j \), its greatest value is \( 1 \) and the minimum value \( \frac{1}{2} \)
for one of the \( 2 \) charginos. In the following section we give the explicit
expression of the branching ratio for the decay \( b\rightarrow s\gamma  \)
and the upper bounds on the mass insertions.

\section{\protect\( BR(b\rightarrow s\gamma )\protect \) Expression and the upper bounds
on the mass insertions \protect\( \delta \protect \).}

The decay \( b\rightarrow s\gamma  \) is very interested because the rare \( B \)
decay represent a good test for new physics beyond SM since they are not affected
appreciably by uncertainties due to long distance effects. 

The branching ratio \cite{10} is \[
BR(b\rightarrow s\gamma )=\frac{BR(B\rightarrow X_{s}\gamma )}{BR(B\rightarrow X_{c}e\overline{\nu _{e}})}=\frac{\Gamma (b\rightarrow s\gamma )}{\Gamma (b\rightarrow ce\overline{\nu _{e}})}\]
 where \( b\rightarrow ce\overline{\nu _{e}} \) is dominant then \( BR(b\rightarrow s\gamma )=\Gamma (b\rightarrow s\gamma )\tau _{_{B}} \).The
explicit expression of \( BR(b\rightarrow s\gamma ) \) obtained from the calculation
exposed in the section \ref{Absg} is: \begin{eqnarray}
BR(b\rightarrow s\gamma ) & =\frac{m^{3}_{b}\alpha \, \tau _{B}}{16\, \pi ^{2}} & \left| \frac{m_{b}\alpha _{s}e_{D}C_{2}(R)}{M^{2}_{D}}\delta _{LL}\, H(X_{\widetilde{g}})\right. \label{bratio}\\
 &  & +\frac{\alpha _{s}e_{D}C_{2}(R)M_{\widetilde{g}}}{M^{2}_{D}}\delta _{LR}\, M_{1}(X_{\widetilde{g}})-\frac{e_{D}\alpha _{w}M_{\chi ^{0}_{j}}\, \sin ^{2}(\theta _{w})\, z_{R\chi _{j}^{0}}}{9\, M^{2}_{D}}\delta _{LR}\, M_{1}(X_{0j})\nonumber\\
 &  & +\frac{e_{D}\alpha _{w}m_{b}\, \sin ^{2}(\theta _{w})\, z_{L\chi ^{0}_{j}}}{18\, M^{2}_{D}\, \cos ^{2}(\theta _{w})}\delta _{LL}\, H(X_{0j})+\frac{m_{b}\alpha _{w}}{M^{2}_{U}}(G(X_{j})+e_{U}\, H(X_{j}))\, \delta _{LL\chi _{j}}\nonumber\\
 &  & +\left. \frac{\alpha _{w}m_{b}m_{t}M_{\chi ^{-}_{j}}}{M^{2}_{w}\, \cos (\beta )\sin (\beta )M_{U}}(\frac{F(X_{j})}{2}-e_{U}M_{1}(X_{j}))\delta _{LR\chi ^{-}_{j}}\right| ^{2}\nonumber
\end{eqnarray}
 By imposing that each individual term, in the above equation, does not exceed
in absolute value the experimental data of \( BR(b\rightarrow s\gamma ) \)
that is \( (1\, -\, 4)10^{-4} \) (which includes the QCD uncertainties following
\cite{4}), we give upper bounds on the different \( \delta  \). 

We have chosen the values for the supersymmetric particles from the experimental
data given in \cite{12}. Moreover, we have imposed the following conditions
: 

\begin{itemize}
\item the average squark mass: \( M_{U}=M_{D}=M_{\widetilde{q}} \).
\item For the neutralino masses: we choose the LSP mass,  with \( M_{\chi _{1}^{0}}\approx \frac{M_{\widetilde{g}}}{6} \)
(GUT relation n)
\item The lightest stop mass: \( M_{\widetilde{t}}\geq M_{\chi _{1}^{0}}+30Gev \) 
\item The chargino mass: \( M_{\chi _{1}^{-}}\approx 2M_{\chi ^{0}_{1}} \) (GUT)
\end{itemize}
Otherwise, we lay down \( X_{\widetilde{g}}=\frac{M^{2}_{\widetilde{g}}}{M_{\widetilde{q}}^{2}} \),
\( X_{0}=\frac{M^{2}_{\chi ^{0}}}{M^{2}_{\tilde{q}}} \), \( X=\frac{M^{2}_{\chi ^{-}}}{M^{2}_{\tilde{q}}} \).
The others experimental data chosen are defined in the following table :

\vspace{0.3cm}
{\centering \begin{tabular}{|c|c|c|c|c|c|c|c|c|}
\hline 
\( \alpha _{s} \)&
\( \alpha  \)&
\( \alpha _{w} \)&
\( \sin ^{2}(\theta _{w}) \)&
\( BR(b\rightarrow s\gamma ) \)&
\( M_{W} \)&
\( m_{b} \)&
\( m_{t} \)&
\( \tau _{B} \)\\
\hline 
0.12&
1/127.9&
\( \alpha /s^{2}_{w} \)&
0.2315&
\( 1\rightarrow 4\, 10^{-4} \) &
80.41&
4.5&
170 &
1.49 \( 10^{-12} \) s\\
\hline 
\end{tabular}\par}
\vspace{0.3cm}

The results are given, in tables \ref{tab:gneut} and \ref{tab:charg}, for
different \( \tan (\beta ) \) values ( i.e. \( \tan (\beta )= \)2, 5, 10,
20 et 40) and for 2 values of \( M_{\tilde{q}} \): \( M_{\tilde{q}}=300\mathrm{GeV} \)
et \( M_{\tilde{q}}=500\mathrm{GeV} \) (the corresponding results are noted
between parenthesis in tables). We have obtained the \( LL \) insertion limits
by imposing \( BR(b\rightarrow s\gamma )=(4)\times 10^{-4} \) and for \( LR \)
insertion \( BR(b\rightarrow s\gamma )=(2)\times 10^{-4} \) to be consistent
with reference \cite{6} in the gluino case.For chargino contribution we take
\( M_{U}=M_{\tilde{q}} \).
\begin{table}
{\centering \begin{tabular}{|c|c|c|c||c|c|c|c|}
\hline 
\( M_{\widetilde{g}} \)&
 \( X_{\tilde{g}} \)&
 \( \delta _{LL} \)&
 \( \delta _{LR} \)&
\( M_{\chi _{1}^{0}} \)&
 \( X_{0} \)&
 \( \delta _{LL}\, z_{L\chi ^{0}_{1}} \)&
 \( \delta _{LR}\, z_{R\chi ^{0}_{1}} \)\\
\hline 
\hline 
\hline 
\( 300 \)&
 \( 1 \)&
 \( 2.96 \)&
 \( 10^{-2} \)&
50&
 \( 3\times 10^{-2} \)&
 \( 7.1 \)&
 \( 0.34 \)\\
&
 \( (0.36) \)&
 \( (8.2) \)&
 \( (2.7\times 10^{-2}) \)&
&
 \( (10^{-2}) \)&
 \( (19.7) \)&
 \( (0.94) \)\\
\hline 
\( 600 \)&
 \( 4 \)&
 \( 9.5 \)&
 \( 1.8\times 10^{-2} \)&
100&
 \( 10^{-2} \)&
 \( 8.4 \)&
 \( 0.25 \)\\
&
 \( (1.44) \)&
 \( (26.4) \)&
 \( (4.9\times 10^{-2}) \)&
&
 \( (4\times 10^{-2}) \)&
 \( (23.25) \)&
 \( (0.7) \)\\
\hline 
\( 800 \)&
 \( 7 \)&
 \( 17.6 \)&
 \( 2.6\times 10^{-2} \)&
130&
 \( 0.19 \)&
 \( 9.8 \)&
 \( 0.23 \)\\
&
 \( (2.56) \)&
 \( (48.8) \)&
 \( (7.2\times 10^{-2}) \)&
&
 \( (7\times 10^{-2)} \)&
 \( (27.3) \)&
 \( (0.64) \) \\
\hline 
\end{tabular}\par}

\caption{\label{tab:gneut}
Limits on the off-diagonal terms \protect\( \delta _{LL}\protect \) and
\protect\( \delta _{LR}\protect \) for down squarks with $M_{\tilde q}=300$GeV
(or $M_{\tilde  q}=500$GeV), coming from gluino and neutralino contributions.}
\end{table}

\begin{figure}
{\par\centering \input{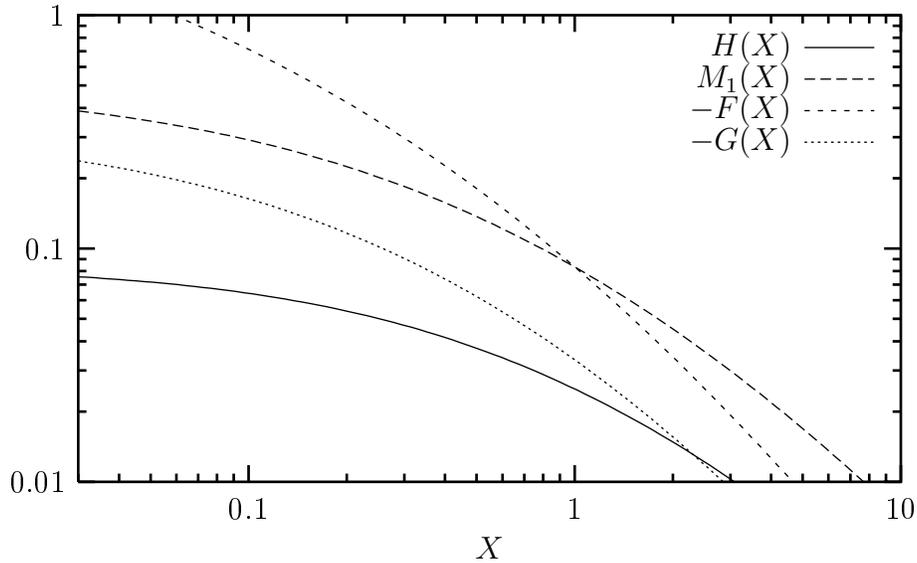}\par}

\caption{\label{functions}The different functions encountered in the evaluation of
the amplitudes, for the available values of \protect\( X=M^{2}_{\mathrm{gaugino}}/M^{2}_{\mathrm{squarks}}\protect \)
on a logarithmic scale.}
\end{figure}

\begin{table}
{\centering \begin{tabular}{|c|c|c|c|c|c|c|c|}
\hline 
\( M_{\chi ^{-}} \)&
 \( X \)&
 \( \delta _{LL\chi }z_{\chi L1} \)&
 \( \delta _{LR\chi }z_{R1} \)&
 \( \delta _{LR\chi }z_{R1} \)&
 \( \delta _{LR\chi }z_{R1} \)&
 \( \delta _{LR\chi }z_{R1} \)&
 \( \delta _{LR\chi }z_{R1} \)\\
&
 \( \tan (\beta )=2 \)&
 \( M_{U=M_{q}} \)&
&
 \( \tan (\beta )=5 \)&
 \( \tan (\beta )=10 \)&
 \( \tan (\beta )=20 \)&
 \( \tan (\beta )=40 \)\\
\hline 
\hline 
\( 100 \)&
 \( 0.1 \)&
 \( 0.57 \)&
 \( 0.08 \)&
 \( 0.04 \)&
 \( 0.02 \)&
 \( 0.01 \)&
 \( 0.0051 \)\\
&
 \( (4\times 10^{-2}) \)&
 \( (1.58) \)&
 \( (0.09) \)&
 \( (0.045) \)&
 \( (0.023) \)&
 \( (0.011) \)&
 \( (0.0058) \)\\
\hline 
\( 200 \)&
 \( 0.44 \)&
 \( 1.23 \)&
 \( 1.6 \)&
 \( 0.76 \)&
 \( 0.39 \)&
 \( 0.2 \)&
 \( 0.099 \)\\
&
 \( (0.16) \)&
 \( (3.41) \)&
 \( (0.2) \)&
 \( (0.095) \)&
 \( (0.049) \)&
 \( (0.025) \)&
 \( (0.012) \)\\
\hline 
\( 300 \)&
 \( 1 \)&
 \( 2.34 \)&
 \( 0.3 \)&
 \( 0.14 \)&
 \( 0.075 \)&
 \( 0.03 \)&
 \( 0.019 \)\\
&
 \( (0.36) \)&
 \( (6.5) \)&
 \( (0.83) \)&
 \( (0.4) \)&
 \( (0.2) \)&
 \( (0.1) \)&
 \( (0.052) \) \\
\hline 
\end{tabular}\par}

\caption{\label{tab:charg}Limits on the off-diagonal terms \protect\( \delta _{LL}\protect \)
and \protect\( \delta _{LR}\protect \) for up squarks with $M_{\tilde q}=300$GeV
(or $M_{\tilde  q}=500$GeV), coming from charginos
contributions.}
\end{table}
From the results obtained we remark that: 
\begin{itemize}
\item because of our use of the mass insertions, we are limited to
\(\delta<1\) for the branching ratio expression (\ref{bratio}) to make
sense. When larger than one, experimental bounds on $\delta$ like
$\delta_{LL}<2.96$ thus really mean that the maximal effect of such
terms (for $\delta_{LL}\sim1$) only contributes a fraction of about
$1/9$th  of the experimental bound;
\item in the gluino case, \( \delta _{LL} \) is more sensitive to the gluino mass
than \( \delta _{LR} \), because this last contribution has an amplitude enhancement
factor of \( M_{\widetilde{g}} \).  Then the dependence on \( M_{\tilde{g}} \)
of the \( H(X_{g}) \) function is partially compensated (see the definitions
in Annex \ref{Annexe} and the plot \ref{functions});
\item for the neutralino, the limit on \( \delta _{LL} \) is less sensitive to the neutralino
mass, thanks to the small values of \( X_{0} \) contained in the same \( H(X_{0}) \)
function. However, even for the upper value for \( z_{L\chi _{j}^{0}}=1 \)
(e.g. for \( j=2 \)), the limit cannot be more competitive than the gluino
limit,  except if \( \frac{M_{\chi ^{0}}}{M_{\widetilde{g}}} \) is smaller
than in the msugra models. The \( \delta _{LR} \) limits decrease weakly with
the enhancement of \( M_{\chi ^{0}} \), due to the additional power of \( M_{\chi ^{0}} \)
in the amplitude and the fact that the function \( M_{1}(X_{0}) \) is fairly
constant for small \( X_{0} \) (see Annex \ref{Annexe} and plot \ref{functions});
\item the chargino contribution is the only one constraining the differences
  between the up squark masses. However the expressions for \( \delta _{LL\chi
    } \) and \( \delta _{LR\chi } \) contain Cabbibo mixing factors. For \(
  \delta _{LR\chi } \), the dominating top contribution allows to extract a
  simple factor of \( V_{ts}V_{tb}\approx \frac{1}{30} \). The first remark
  above then applies for interpreting our mass insertions results, once
  $\delta_{LL\chi}>0.03$. Nevertheless, we obtain a limit that quickly becomes
  more constraining with increasing \( \tan (\beta ) \). For
  $\delta_{LL\chi}$, the factorization of CKM elements cannot be so easily
  justified: if for instance there is large mixing between the squarks 2 and
  3, the leading CKM factors are of order one, and the expression
  (\ref{delta_LLXj}) for $\delta_{LL\chi}$ with up squarks becomes the same as
  $\delta_{LL}$ for down squarks in (\ref{delta_LL}). The sensitivity on
  $M_{\tilde q}$ of the bounds is about the same factor of 3 for the chargino
  $\delta_{LL\chi}$ as for the neutralino $\delta_{LL}$ and $\delta_{LR}$
  while the chargino $\delta_{LR\chi}$ is less sensitive for small chargino
  masses;
\item the limits in tables (1) and (2) can easily be generalized for different
  values of squark and gaugino masses: they are inversely proportional to the
  functions given in Annex \ref{Annexe} and plotted in figure \ref{functions},
\item finally, if following \cite{13}, we use the latest CLEO data \cite{14} and
substract the SM contribution \cite{15}, we obtain a tighter bound, in the
absence of cancellations between the various contributions that was assumed
throughout this paper. The various results in the tables are then reduced by
a factor of about 20. 
\end{itemize}

\section{Conclusion}

We have given explicit expressions for the amplitudes associated to the supersymmetric
contributions to the decay \( b\rightarrow s\gamma  \) in the context of supersymmetric
extensions of SM with non-universal soft SUSY breaking terms. The model independent
parameterization which we have chosen  is the mass insertion approximation.
From the FCNC experimental data, we have derived upper bounds on the different
\( \delta  \)'s. The contribution from the chargino and neutralino exchanges
are less sensitive to the gaugino mass than the gluino contribution.

\section{\label{Annexe}Annex}

\begin{eqnarray*}
H(X) & = & \frac{-1+9X+9X^{2}-17X^{3}+6X^{2}(X\ln X+3\ln X)}{12(X-1)^{5}}\\
M_{1}(X) & = & \frac{1+4X-5X^{2}+2X(X+2)\ln (X)}{8(X-1)^{4}}=L(X)/2\\
F(X) & = & \frac{5-4X-X^{2}+2\ln (X)+4X\ln (X)}{2(X-1)^{^{4}}}\\
G(X) & = & \frac{1+9X-9X^{2}-X^{3}+6X(1+X)\ln (X)}{3(X-1)^{5}}
\end{eqnarray*}

\subsection*{Note added}
This work was supported in part by the GdR Supersymetrie of the French
CNRS.

While this work was being refereed, a general study of \( B\rightarrow X_s
\gamma \) appeared\cite{16} with interesting results on the interferences
between various contributions, including the ones presented here. For the
parameters studied there ($\mu=300$GeV, $M_{\tilde q}=500$GeV,
$M_2=100\rightarrow 230$GeV), we agree that there is no constrain on
up squarks from table \ref{tab:charg}: the light chargino is a gaugino with
$z_{R1}\ll 1$ and the heavy chargino gives the last line in parenthesis, with
$z_{R2}=1$. Taking the Cabbibo factors into account, the strongest bound on
the off-diagonal element $\delta_{u,LR,23}$ is just about one for
$\tan\beta=50$, which means no contraint in the mass insertion logics.

\end{document}